%% file: TAC_CLF_ConvexOptim.tex
\documentclass[journal]{IEEEtran}

\IEEEoverridecommandlockouts

\usepackage{makeidx}         
\usepackage{graphicx}        
                             
\usepackage{multicol}

\usepackage{amsmath,amssymb}
\usepackage{euscript}
\usepackage{subfig}
\usepackage{url}
\usepackage{mathrsfs}
\usepackage{array}
\usepackage{wrapfig}

\usepackage{epsfig,float,alltt}
\usepackage{psfrag,xr}
\usepackage{pdfpages}

\usepackage[T1]{fontenc}

\usepackage{multirow}

\newtheorem{remark}{\bfseries Remark}

\renewcommand{\epsilon}{\varepsilon}

\newcommand{\q}{q}
\newcommand{\dq}{\dot{q}}

\newcommand{\etavar}{\eta}

\newcommand{\Ve}{V_{\epsilon}}

\newcommand{\mabel}{MABEL}

\newcommand{\mLS}{{\rm{mLS}}}
\newcommand{\mLA}{{\rm{mLA}}}
\newcommand{\Tor}{{\rm{Tor}}}
\newcommand{\LA}{{\rm{LA}}}
\newcommand{\LS}{{\rm{LS}}}
\newcommand{\stlbl}{{\rm{st}}} 
\newcommand{\swlbl}{{\rm{sw}}} 

\begin{document}

 \title{Torque Saturation in Bipedal Robotic Walking through Control Lyapunov Function Based Quadratic Programs} 

\author{Kevin Galloway, Koushil Sreenath, Aaron D. Ames, and J. W. Grizzle
\thanks{The work of A.~D. Ames is supported by NSF grants CNS-0953823 and CNS-1136104, NHARP project 000512-0184-2009 and NASA contract NNX12AB58G, K. ~Galloway is supported by DARPA Contract W91CRB-11-1-0002, and J. ~Grizzle is supported in part by DARPA and by NSF grant ECS-909300.}
\thanks{A. D. Ames is with the Department of Mechanical Engineering, Texas A\&M University, College Station, TX 77843, email: {\tt\small aames@tamu.edu}.}
\thanks{K. Galloway and J. W. Grizzle are with the Control Systems Laboratory, Electrical Engineering and Computer Science Department, University of Michigan, Ann Arbor, MI 48109, email: {\tt\{kevinsg,grizzle\}@umich.edu}.}
\thanks{K. Sreenath is with the GRASP Laboratory, Department of Mechanical Engineering and Applied Mechanics, University of Pennsylvania, Philadelphia, PA 19104, email: {\tt koushils@seas.upenn.edu}.}
}

\maketitle

\input{sections/abstract-intro}

\input{sections/CLF_revisited}

\input{sections/CLF_convex_optim}

\input{sections/results}

\input{sections/conclusion}

\bibliographystyle{plain}
\bibliography{bibdata,biped,Ames_ref,clf_ref}

\end{document}

%% file: sections/abstract-intro.tex
\begin{abstract}
This paper presents a novel method for directly incorporating user-defined control input saturations into the calculation of a control Lyapunov function (CLF)-based walking controller for a biped robot. Previous work by the authors has demonstrated the effectiveness of CLF controllers for stabilizing periodic gaits for biped walkers \cite{AmGaGrSr:TAC12}, and the current work expands on those results by providing a more effective means for handling control saturations. The new approach, based on a convex optimization routine running at a 1 kHz control update rate, is useful not only for handling torque saturations but also for incorporating a whole family of user-defined constraints into the online computation of a CLF controller. The paper concludes with an experimental implementation of the main results on the bipedal robot \mabel.   
\end{abstract}

\section{Introduction}
Biped locomotion presents an interesting control challenge, especially since the dynamic models are typically hybrid and underactuated. The method of Hybrid Zero Dynamics (HZD) \cite{WEGRKO03,WEBUGR04,WGCCM07} has provided a rigorous and intuitive method for implementing periodic walking gaits in such robotic systems, by driving the system to a lower-dimensional zero dynamics manifold on which the walking gait exists as an exponentially stable periodic orbit. Typical experimental implementation of the HZD method has relied on input-output linearization with PD control to drive the system to the zero dynamics manifold \cite{KOPAPOG210}, but recent work by the authors has demonstrated that control Lyapunov function (CLF)-based controllers can be used to effectively implement stable walking with smoother control torques, both in simulation and in experimental contexts \cite{AmGaGr:CDC12,AmGaGrSr:TAC12}.

CLF's provide a means for not only guaranteeing exponential stability of a system but also providing an explicit bound on the rate of convergence. In the case of hybrid systems (such as biped robots with impulsive foot-ground impact), a stronger convergence property is required, and therefore we turn to rapidly exponentially stabilizing control Lyapunov functions (RES-CLF). This type of CLF, which will be reviewed in more detail in Section \ref{sec:CLF_revisited}, incorporates an additional tuning parameter which allows the user to directly control the rate of exponential convergence. 
The work in \cite{AmGaGrSr:TAC12} established the key theoretical properties of CLF controllers in a hybrid context, and also presented a description of the sucessful experimental implementation of a CLF-based controller on the robotic testbed MABEL. In fact, analysis of the experimental data showed that the control torques generated by the CLF controller were much smoother and less noisy than the control torques generated by a comparable PD controller. However, it was also noted that the user-defined control saturations were active throughout a large portion of the walking experiment, and that these saturations had a significant impact on the actual performance of the CLF controller as compared to the predicted performance based on theoretical bounds. In this context the hard saturation bounds were ``blindly'' applied to the calculated CLF control torques, with no consideration of the potential effect on the CLF controller performance. 

The main contribution of this paper is to present an alternative method of controller implementation that not only preserves the desirable performance characteristics promised by the CLF theory, but also respects the user-defined saturation bounds on the inputs. To the authors' knowledge, CLF controllers that respect saturations have not been addressed before, and this paper provides a constructive technique for doing so, while also presenting an experimental implementation on an embedded hard real-time system with a high control rate of 1 kHz. 

The paper proceeds as follows. In Section \ref{sec:CLF_revisited}, we state the dynamics of the relevant model and review the results on CLF-based control of biped robots from \cite{AmGaGr:CDC12} and \cite{AmGaGrSr:TAC12}. Section \ref{sec:saturationEffects} discusses the adverse effects of user-specified control input saturations on the CLF controller, providing the motivation for Section \ref{sec:CLF_convex_optim} which introduces a new method for using quadratic programming to appropriately handle torque saturation constraints for the CLF controllers. Section \ref{sec:results} presents simulation and experimental results, and we conclude in Section \ref{sec:conclusion} with a summary and discussion of future work.

%% file: sections/CLF_revisited.tex
\section{Control Lyapunov Functions for Hybrid Systems Revisted}
\label{sec:CLF_revisited}
In this section we introduce the model for a biped robot and review the recent innovations introduced in \cite{AmGaGr:CDC12} and \cite{AmGaGrSr:TAC12} for using control Lyapunov functions to control such systems. The dynamics for a biped robot (such as \mabel, the robot described in Section \ref{sec:results}) can be derived by the standard method of Lagrange and take the form
\begin{align}
\label{eqn:robotDynamicsDCGB}
	D(q)\ddot{q} + C(q, \dot{q})\dot{q} + G(q) = B(q)u,
\end{align}
where $q \in \mathscr{Q}$ is the robot configuration variable and $u$ is the motor control torques. The configuration $q$ for the particular case of \mabel\ is described in \cite{KOPAPOG210} and depicted in Figure \ref{fig:MABELcoordinates}. Reformulating the dynamics \eqref{eqn:robotDynamicsDCGB} as
\begin{eqnarray}
\label{eqn:fgsystemRepeated}
\left[ \begin{array}{c} \dot{q} \\ \ddot{q} \end{array} \right] = f(q,\dot{q}) + g(q,\dot{q}) u,
\end{eqnarray}
we also define output functions of the form
\begin{align}
\label{eqn:robotOutputFcns}
	y(q) := H_{0} q - y_{d}(\theta(q)),
\end{align}
where $\theta(q)$ is a strictly monotonic function of the configuration variable $q$. The method of Hybrid Zero Dynamics (HZD) aims to drive these output functions (and their first derivatives) to zero, thereby imposing ``virtual constraints'' such that the system evolves on the lower-dimensional zero dynamics manifold, given by
\begin{align}
\label{eqn:zeroManifold}
Z=\{ (q,\dot{q}) \in T\mathscr{Q}~|~ y(\q) = 0, \:\: L_{f}y(q,\dot{q}) = 0\}.	
\end{align} 
If $y(q)$ has relative degree 2, then the second derivative takes the form
\begin{equation}
\label{eqn:yddot}
\ddot{y} = L^{2}_{f}y(\q,\dq) + L_{g}L_{f} y(\q,\dq) u,
\end{equation}
where the decoupling matrix $L_{g}L_{f} y(\q,\dq)$ is invertible due to the vector relative degree assumption. Then defining 
\begin{align}
\label{eqn:ustar}
	u^{*}(q,\dot{q}) := -(L_{g}L_{f} y(q,\dot{q}))^{-1} L^2_{f} y(q,\dot{q}),
\end{align}
and applying a pre-control law of the form 
\begin{align}
\label{eqn:preControl}
	u(q,\dot{q}) = u^{*}(q,\dot{q}) + \mu
\end{align}
or
\begin{align}
\label{eqn:expControl}
	u(q,\dot{q}) = u^{*}(q,\dot{q}) + (L_{g}L_{f} y(q,\dot{q}))^{-1}\mu
\end{align}
renders $Z$ invariant (provided $\mu$ vanishes on $Z$). (Note that $u^{*}(q,\dot{q})$ is a feed-forward term representing the torque required to remain on $Z$.)

Under these assumptions, the dynamics \eqref{eqn:fgsystemRepeated} can be decomposed into zero dynamics states $z \in Z$ and transverse variables $\eta = \begin{bmatrix} y   &  \dot{y} \end{bmatrix}$. (See \cite{WGCCM07,ISI89} for details.) Under a pre-control law of the form \eqref{eqn:preControl} or \eqref{eqn:expControl}, the closed-loop dynamics in terms of $(\eta,z)$ take the form
\begin{align}
\label{eqn:fgBarDynamics}
\dot{\eta} &= \bar{f}(\eta, z)+ \bar{g}(\eta, z)\mu \\
\dot{z} & = p(\eta,z). 
\end{align}
For the work presented here, we will use the pre-control law \eqref{eqn:expControl} so that $\bar{f}(\eta, z) = F\eta$ and $\bar{g}(\eta, z) = G$, where 
\begin{align}
\label{eqn:FGdefns}
F = \left[ \begin{array}{cc} 0  & I \\ 0 & 0 \end{array} \right], \qquad
G =  \left[ \begin{array}{c} 0 \\ I \end{array} \right].
\end{align}

The most common approach to controlling the transverse variables (i.e. driving $\eta$ to zero) relies on input-output linearization with PD control, using \eqref{eqn:expControl} with $\mu = \left[ \begin{array}{cc} -\frac{1}{\epsilon^{2}}K_P & -\frac{1}{\epsilon}K_D \end{array} \right]\eta$, where $K_P$ and $K_D$ are diagonal matrices chosen such that 
\begin{equation}
\label{eqn:Amatrix}
A:= \begin{bmatrix} 0 & I \\ -K_P & - K_D  \end{bmatrix}
\end{equation}
is Hurwitz. Recently, a new method based on control Lyapunov functions has been introduced in \cite{AmGaGr:CDC12,AmGaGrSr:TAC12}, which provides an alternative method for controlling the transverse variables. That method can be summarized as follows.

A function $\Ve(\eta)$ is a \textit{rapidly exponentially stabilizing control Lyapunov function (RES-CLF)} for the system \eqref{eqn:fgBarDynamics} if there exist positive constants $c_1, c_2, c_3 > 0$ such that for all $0 < \epsilon < 1$ and all states $(\eta,z)$ it holds that
\begin{align}
\label{eq:Vineqepsilon}
 & c_1 \| \eta \|^2 \leq V_{\epsilon}(\eta) \leq \dfrac{c_2}{\epsilon^2} \|\eta \| ^2  \\
 \label{eq:Vinfepsilon}
&\inf_{\mu \in U} \left[ L_{\bar{f}} V_{\epsilon}(\eta,z) + L_{\bar{g}} V_{\epsilon}(\eta,z) \mu + \frac{c_3}{\epsilon} V_{\epsilon}(\eta) \right] \leq 0.
\end{align}
One way to generate a RES-CLF $\Ve(\eta)$ is to first solve the Lyapunov equation $A^T P + P A = -Q$ for $P$ (where $A$ is given by \eqref{eqn:Amatrix} and $Q$ is any symmetric positive-definite matrix), and then define
\begin{equation}
\label{eqn:VUnscaledCoords}
\Ve(\eta) = \eta^{T}\begin{bmatrix} \frac{1}{\epsilon}I & 0 \\ 0 & I  \end{bmatrix} P \begin{bmatrix} \frac{1}{\epsilon}I & 0 \\ 0 & I  \end{bmatrix} \eta =: \eta^{T} P_\epsilon \eta,
\end{equation}
for which we have 
\begin{align}
\label{eqn:LfVLgVdefns2}
L_{\bar{f}} V_{\epsilon}(\etavar,z)& = \etavar^T (F^T P_{\epsilon} + P_{\epsilon} F) \etavar,  \nonumber\\
L_{\bar{g}} V_{\epsilon}(\etavar,z)& = 2 \etavar^T P_{\epsilon} G.
\end{align}

Associated with a RES-CLF is the set of all $\mu$ for which \eqref{eq:Vinfepsilon} is satisfied, 
{\small
$$
K_{\epsilon}(\eta,z) = \{ \mu \in U : L_{\bar{f}} \Ve(\eta,z) + L_{\bar{g}} \Ve(\eta,z) \mu + \frac{c_3}{\epsilon} \Ve(\eta) \leq 0\},
$$}and one can show that for any Lipschitz continuous feedback control law $\mu_{\epsilon}(\eta,z) \in K_{\epsilon}(\eta,z)$, it holds that
\begin{eqnarray}
\label{eqn:xboundV}
\| \eta(t) \| \leq \frac{1}{\epsilon}\sqrt{\frac{c_2}{c_1}} e^{-\frac{c_3}{2 \epsilon} t} \| \eta(0) \|,
\end{eqnarray}
i.e. the rate of exponential convergence to the zero dynamics manifold can be directly controlled with the constant $\epsilon$ through $\frac{c_3}{\epsilon}$. There are various methods for finding a feedback control law $\mu_{\epsilon}(\eta,z) \in K_{\epsilon}(\eta,z)$; in practical applications, it is often important to select the control law of minimum norm. If we let $c_3 = \frac{\lambda_{\min}(Q)}{\lambda_{\max}(P)}$ and define
\begin{align}
\label{eqn:psiDefnsForExperiment}
\psi_{0,\epsilon}(\eta,z) & =  L_{\bar{f}} V_{\epsilon}(\etavar,z) + \frac{c_3}{\epsilon} V_{\epsilon}(\eta,z) \nonumber\\
\psi_{1,\epsilon}(\eta,z) & = L_{\bar{g}} V_{\epsilon}(\etavar,z)^{T},
\end{align}
then this pointwise min-norm control law \cite{FK:Book} can be explicitly formulated as 
\begin{align}
\label{eqn:minnorm}
	\mu_{\epsilon}(\eta,z) = \left\{ \begin{array}{lcr}  - \frac{\psi_{0,\epsilon}(\eta,z)\psi_{1,\epsilon}(\eta,z)}{\psi_{1,\epsilon}(\eta,z)^T \psi_{1,\epsilon}(\eta,z)} & \mathrm{if} & \psi_{0,\epsilon}(\eta,z) > 0 \\
0 & \mathrm{if} & \psi_{0,\epsilon}(\eta,z) \leq 0,   \end{array} \right\}
\end{align}
wherein we can take $\mu = \mu_{\epsilon}$ in \eqref{eqn:expControl}. 

\section{Adverse effects of torque saturation on the CLF controller}
\label{sec:saturationEffects}
\begin{figure}
		\centering
		\psfrag{Time (s)}[Bc][Bc][1.5]{Time (s)}
		
		\psfrag{umLA (Nm)}[Bc][Bc][2]{$u_\mLA$ {\scriptsize (Nm)}}
		\psfrag{umLS (Nm)}[Bc][Bc][2]{$u_\mLS$ {\scriptsize (Nm)}}
		\psfrag{Stance}[Bc][Bc][1]{Stance}
		\psfrag{Swing}[Bc][Bc][1]{Swing}
\resizebox{1.0\linewidth}{!}{\includegraphics{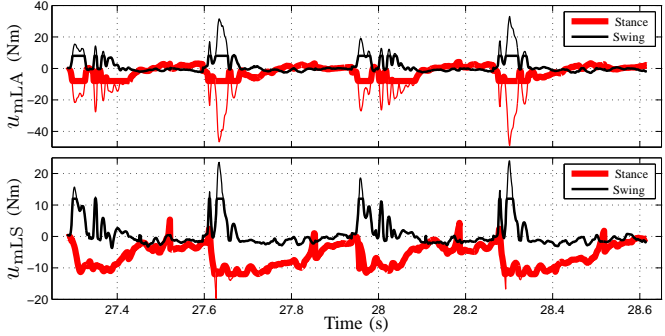}}
		\caption{Motor torques (from the MABEL experiment described in \cite{AmGaGrSr:TAC12}) for the stance and swing legs for $4$ consecutive steps of walking with the minimum-norm CLF controller given in \eqref{eqn:minnorm}.  The thicker plots indicate the experimental (saturated) torques, while the thinner plots are the raw (unsaturated) torques computed by the CLF controller.  For the leg angle motor (top graph), the raw (desired) control torque is at times more than $400$\% of the (actually implemented) saturated value. Moreover, this occurs over a significant duration of the step.}
		\label{fig:CLF_minnorm_torques}
	\end{figure}

The approach described in Section \ref{sec:CLF_revisited} was successfully implemented on the robotic testbed \mabel, producing a stable walking gait with motor torques which were smoother than a comparable approach based on the PD control method. (See \cite{AmGaGrSr:TAC12} for a description of the experiment and a reference to the online video.) However, analysis of the experimental data reveals that the user-imposed saturations on the control torque inputs were active throughout much of the experiment (see Figure \ref{fig:CLF_minnorm_torques}) and significantly affected the implementation of the CLF control method. Though necessary to prevent unsafe or damaging motions, these saturation constraints were not applied in a manner that appropriately preserved the qualities of the CLF controller, and therefore the nominal bounds given by \eqref{eq:Vinfepsilon} and \eqref{eqn:xboundV} were frequently violated.

Saturation bounds for control inputs are typically imposed by the user either as a measure of safety or out of physical necessity, as in the case when current draw must be limited due to a dying battery. When the calculated ideal control effort frequently exceeds the prescribed saturation bounds, the controller performance is degraded and theoretical performance measures may be violated, as in the experiment described above. More importantly, when a control input is saturated, the system runs in open-loop and is no longer able to respond to increasing errors in tracking, often leading to eventual failure. 

Designing controllers which respect such saturation bounds is important (especially in the experimental context), and therefore a variety of approaches have been developed, such as quasi-linear control \cite{ChEuGoKaMe:QLC2011}, which offers one solution for a special class of systems.  The main objective of the current work is to present a method for implementing CLF-based controllers for a general class of nonlinear systems in a manner which respects the user-specified input saturations.

%% file: sections/CLF_convex_optim.tex
\section{Formulating the CLF Min-Norm controller as a Convex Optimization}
\label{sec:CLF_convex_optim}
		To design such a controller, we proceed by recognizing that the pointwise min-norm controller in \eqref{eqn:minnorm} can be equivalently expressed as a convex optimization problem formulated as
\begin{equation} \label{eq:minnorm_convexoptim}
	\begin{aligned}
	& \underset{\mu}{\min} & & \mu^T \mu \\
	& \text{s.t.} & & \psi_{0,\epsilon}(\eta,z) + \psi_{1,\epsilon}(\eta,z)~\mu \le 0.
	\end{aligned}
\end{equation}
The inequality constraint above enforces the bound on the time-derivative of the CLF given by \eqref{eq:Vinfepsilon}, which can be equivalently expressed as $\dot{V}_\epsilon(\eta) \le -c_3/\epsilon ~ V_\epsilon(\eta)$.  The solution of this convex optimization problem is then exactly the controller specified in \eqref{eqn:minnorm}.

    Once we have expressed the pointwise min-norm controller as a convex optimization problem, we can introduce saturation bounds on the control input in the form of additional constraints for the convex optimization problem.  However, for these additional constraints to be satisfied, we first need to relax the bound on the time-derivative of the CLF.  We do this by requiring $\dot{V}_\epsilon(\eta) \le -c_3/\epsilon ~ V_\epsilon(\eta) + d_1$, where $d_1$ is typically a small positive quantity. We then introduce input saturations in the form of (soft) bounds, such that the control input $u$ in \eqref{eqn:expControl} satisfies $u_{min} - d_2 \le u \le u_{max} + d_3$, with $d_2, d_3$ typically small positive quantities.  The new optimization problem is formulated as
\begin{equation} \label{eqn:cvxWithSoftSaturations}
	\begin{aligned}
	& \underset{\mu,d_1,d_2,d_3}{\min} & & \mu^T \mu + p_1~d_1^2 + p_2~(d_2^2 + d_3^2)\\
	& \text{s.t.} & & \psi_{0,\epsilon}(\eta,z) + \psi_{1,\epsilon}(\eta,z)~\mu \le d_1, \\
	& & & (L_gL_f y(q,\dot{q}))^{-1} ~ \mu \ge (u_{min} - u^*) - d_2, \\
	& & & (L_gL_f y(q,\dot{q}))^{-1} ~ \mu \le (u_{max} - u^*) + d_3,
	\end{aligned}
\end{equation}
where $p_1,p_2$ are large positive numbers that represent the penalty of relaxing the inequality constraints and $u^*$ is defined by \eqref{eqn:ustar}.  Typically, we choose $p_2 > p_1$ to ensure that violation of the saturation bound on control input is penalized far more severely than violation of the bound on the time-derivative of the CLF.  Note that the constraints on the control inputs enforce $u_{min}-d_2 \le u^*+v \le u_{max}+d_3$, where $v = (L_gL_f y(q,\dot{q}))^{-1}\mu$, i.e., the bound is on the total control input.

    The above formulation provides the control designer with parameters to trade off violation of the bound on the time-derivative of the CLF with that of the saturation bound on the control input.  However, in most cases the bounds on the inputs appear as hard saturation bounds which cannot be relaxed.  In this case, the optimization problem can be redefined to perfectly satisfy torque bounds as,
\begin{equation}
\label{eqn:cvxWithHardSaturations}
	\begin{aligned}
	& \underset{\mu,d_1}{\min} & & \mu^T \mu + p_1~d_1^2 \\
	& \text{s.t.} & & \psi_{0,\epsilon}(\eta,z) + \psi_{1,\epsilon}(\eta,z)~\mu \le d_1, \\
	& & & (L_gL_f y(q,\dot{q}))^{-1} ~ \mu \ge (u_{min} - u^*), \\
	& & & (L_gL_f y(q,\dot{q}))^{-1} ~ \mu \le (u_{max} - u^*).
	\end{aligned}
\end{equation}

\begin{remark}
		Note that in both \eqref{eqn:cvxWithSoftSaturations} and \eqref{eqn:cvxWithHardSaturations} we have depicted $u_{min}$ and $u_{max}$ as constants.  However, since the convex optimization problem is to be solved at every instant in time, these values can be specified as functions of time or system state, leading to \emph{dynamic torque saturation}.  As will be discussed in the next section, for periodic motions (such as bipedal walking), this provides the option to define a control bound that varies according to the system state location along the periodic orbit. 
\end{remark}

%% file: sections/results.tex
\section{Simulation and Experimental Results}
\label{sec:results}

\begin{figure}
\centering
\subfloat[][Coordinates]{
			\psfrag{a}{\scriptsize$-q_{{}_{\Tor}}$}
			\psfrag{b}{\scriptsize$q_{{}_{\LA}}$}
			\psfrag{c}{\scriptsize$q_{{}_{\LS}}$}
			\psfrag{v}{\scriptsize Virtual}
			\psfrag{w}{\scriptsize Compliant Leg}
			\psfrag{d}{\scriptsize $-\theta$}
\resizebox{0.35\linewidth}{!}{\includegraphics{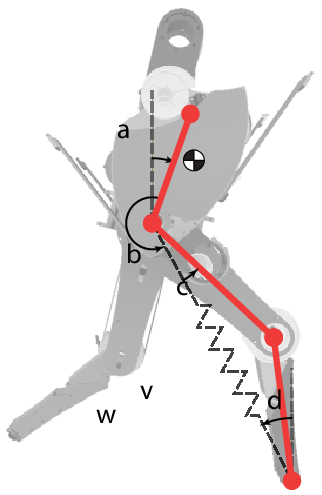}}
\label{fig:MABELcoordinates}}
\subfloat[][MABEL experimental setup]{
\resizebox{0.6\linewidth}{!}{\includegraphics{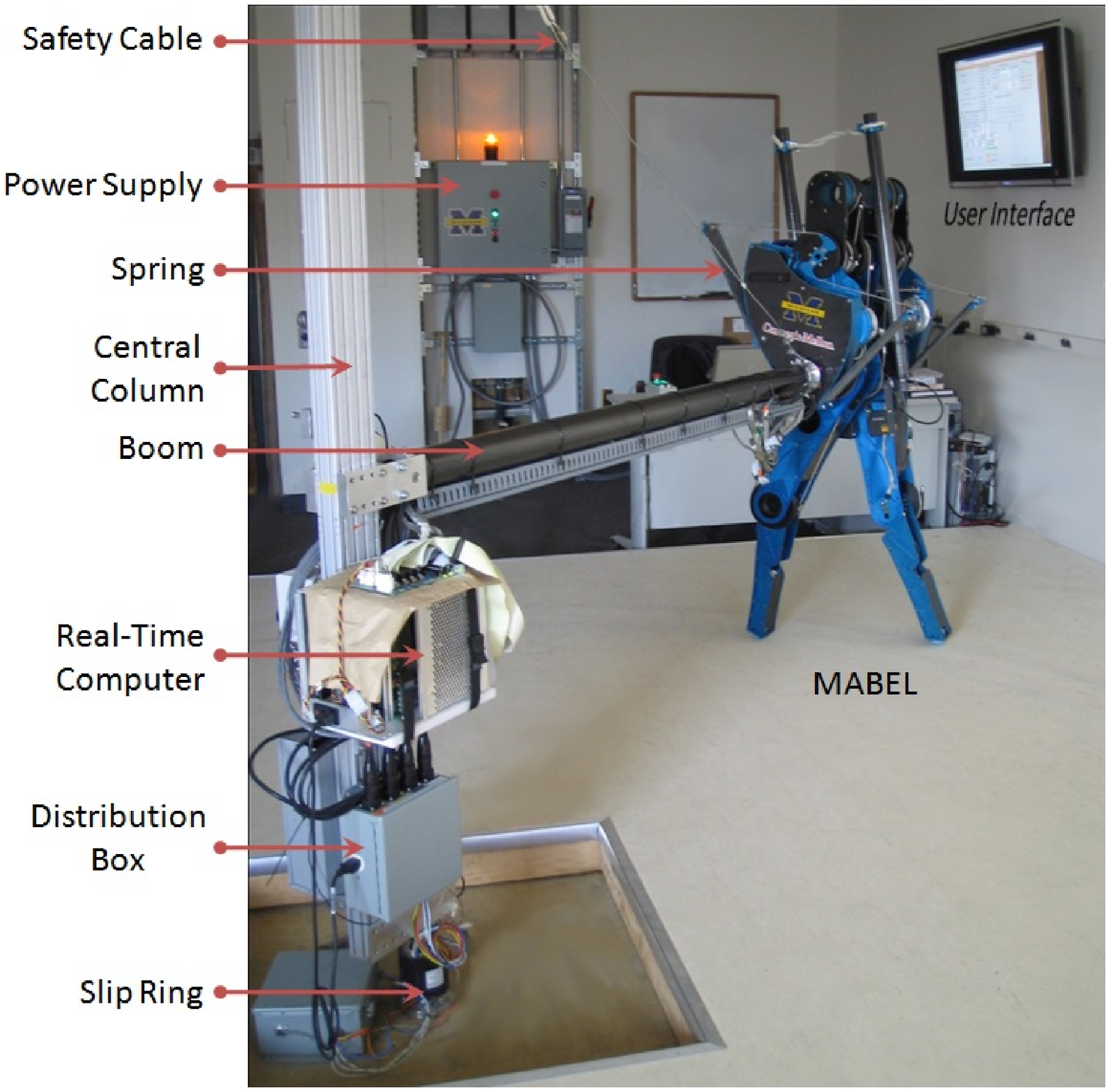}}
\label{fig:MABEL}}
\caption{Experimental setup for bipedal robot MABEL and associated coordinates. (From \cite{KOPAPOG210}.)}
\label{fig:MABELfigs}
\end{figure}

	In this section we present both numerical simulation and experimental results to validate the performance of the control methods described in Section \ref{sec:CLF_convex_optim}. Since experimental testing on \mabel\ was the ultimate goal, the numerical studies were conducted first on a simple model of \mabel, followed by simulations on a complex model of \mabel\, developed in \cite{PaSrHuGr2011}, which closely replicates the experimental setup. This latter model includes a compliant ground model as well as a model that allows for stretch in the cables between the transmission pulleys. \mabel\ is a 5-link bipedal robot with point feet and series-compliant actuation for improved agility and energy efficiency. The experimental setup has been described previously in \cite{KOPAPOG210} and is illustrated in Figure \ref{fig:MABELfigs}. For the simulations and experiments described here, the four output functions in \eqref{eqn:robotOutputFcns} were defined by the absolute pitch angle of the torso, the leg angle (\textit{LA}) for the swing leg, and the appropriately scaled leg-shape motor position (\textit{mLS}) for the swing and stance legs. The four control inputs are the leg-angle motor torque ($u_{\mLA_\stlbl}, u_{\mLA_\swlbl}$) and leg-shape motor torque ($u_{\mLS_\stlbl}, u_{\mLS_\swlbl}$) for the stance and swing legs respectively.

\subsection{Numerical simulation}
\label{sec:simulation}
	The numerical simulation results presented here employ the CLF controller with hard input saturation, as in \eqref{eqn:cvxWithHardSaturations}.  We consider four separate cases with different control saturation bounds, given by
	\begin{eqnarray} \label{eq:simulation_cases}
		A &:&
		\begin{cases}
			\begin{aligned}
				\begin{bmatrix}
					-8 \\ -12 \\ -8 \\ -12
				\end{bmatrix} \le
				\begin{bmatrix}
					u_{\mLA_\stlbl} \\
					u_{\mLS_\stlbl} \\
					u_{\mLA_\swlbl} \\
					u_{\mLS_\swlbl}
				\end{bmatrix} \le
				\begin{bmatrix}
					8 \\ 12 \\ 8 \\ 12
				\end{bmatrix}

			\end{aligned}
		\end{cases} \nonumber \\
		B &:&
		\begin{cases}
			\begin{aligned}
				\begin{bmatrix}
					-5 \\ -8 \\ -2 \\ -2
				\end{bmatrix} \le
				\begin{bmatrix}
					u_{\mLA_\stlbl} \\
					u_{\mLS_\stlbl} \\
					u_{\mLA_\swlbl} \\
					u_{\mLS_\swlbl}
				\end{bmatrix} \le
				\begin{bmatrix}
					4 \\ 4 \\ 4 \\ 4
				\end{bmatrix}

			\end{aligned}
		\end{cases} \nonumber \\
		C &:&
		\begin{cases}
			\begin{aligned}
				\begin{bmatrix}
					-4 \\ -8 \\ -2 \\ -2
				\end{bmatrix} \le
				\begin{bmatrix}
					u_{\mLA_\stlbl} \\
					u_{\mLS_\stlbl} \\
					u_{\mLA_\swlbl} \\
					u_{\mLS_\swlbl}
				\end{bmatrix} \le
				\begin{bmatrix}
					1 \\ 1 \\ 1 \\ 1
				\end{bmatrix}

		\end{aligned}
		\end{cases} \nonumber \\
		D &:&
		\begin{cases}
			\begin{aligned}
				u^*(\theta) + 
				\begin{bmatrix}
					-4 \\ -7 \\ -1 \\ -1
				\end{bmatrix} \le
				\begin{bmatrix}
					u_{\mLA_\stlbl} \\
					u_{\mLS_\stlbl} \\
					u_{\mLA_\swlbl} \\
					u_{\mLS_\swlbl}
				\end{bmatrix} \le
				u^*(\theta) + 
				\begin{bmatrix}
					4 \\ 7 \\ 1 \\ 1
				\end{bmatrix}

			\end{aligned}
		\end{cases} \nonumber
		 \nonumber
	\end{eqnarray}
	where $u^*(\theta)$ is the nominal value of \eqref{eqn:ustar} along the periodic orbit, regressed as a 5th-order Bezier polynomial of $\theta(q)$. Note that in case D, the bounds are specified dynamically as a function of the state of the robot, resulting in dynamic torque saturation.  
	
\begin{figure}
		\centering
		\psfrag{Time (s)}[Bc][Bc][1.5]{Time (s)}
		\psfrag{V}[Bc][Bc][2.5]{$V_\epsilon$}
		\psfrag{dV}[Bc][Bc][2.5]{$\dot{V}_\epsilon$}
		\resizebox{1.0\linewidth}{!}{\includegraphics{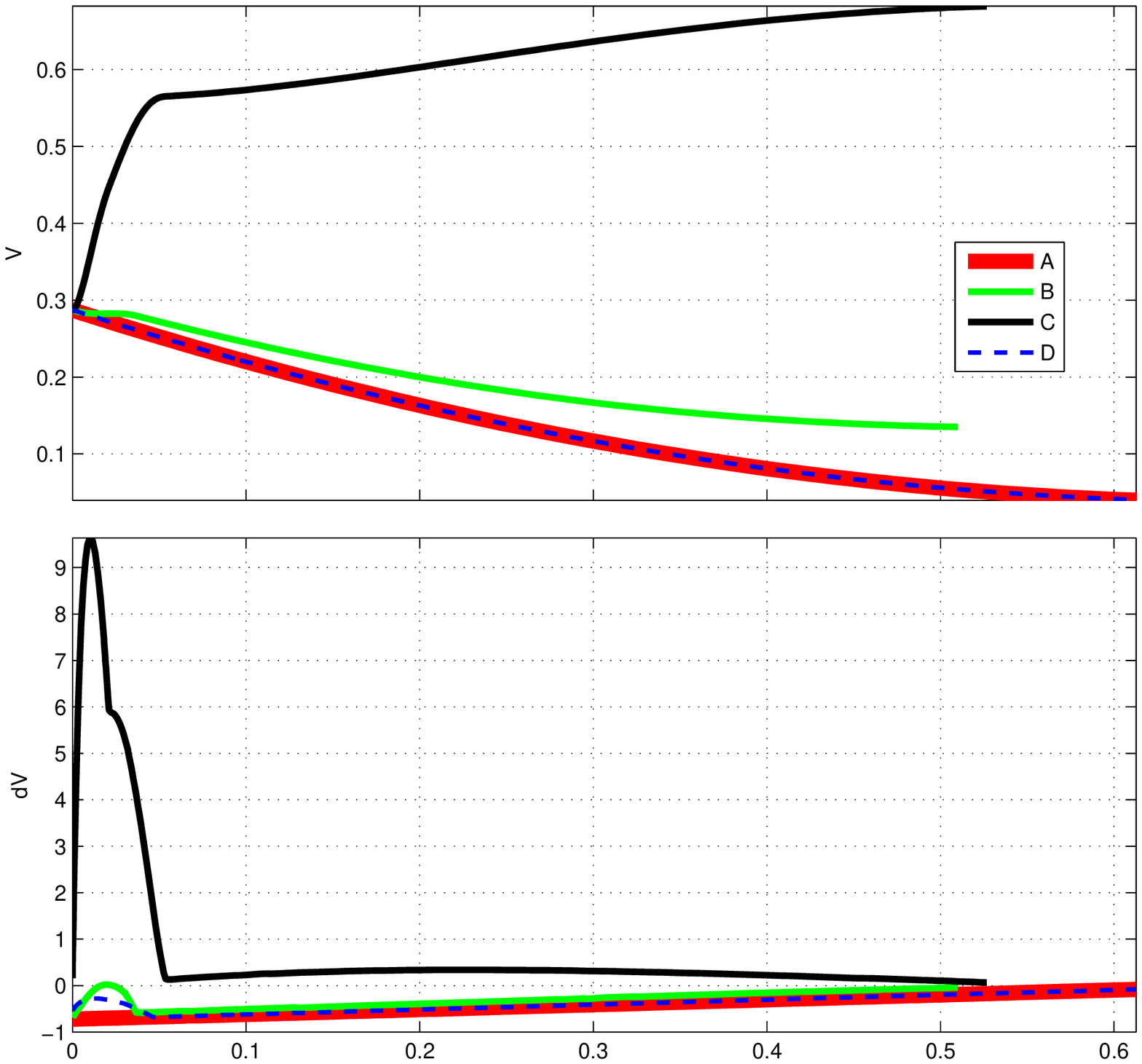}}
		\caption{Lyapunov function $\Ve$ and its derivative for the numerical simulations described in Section \ref{sec:simulation}. The figures depict the results for four different cases of input saturation bounds.  As can be observed in the plot of $\Ve$, Case C leads to instability in the walking gait.}
		\label{fig:sim_V_dV}
	\end{figure}

\begin{figure*}
		\centering
		\subfloat[][]{
			\psfrag{Time (s)}[Bc][Bc][1.5]{Time (s)}
			\psfrag{Control Input}[Bc][Bc][2]{Control Input}
			\psfrag{umLA-st (Nm)}[Bc][Bc][2]{$u_{\mLA_\stlbl}$ {\scriptsize (Nm)}}
			\psfrag{umLA-sw (Nm)}[Bc][Bc][2]{$u_{\mLA_\swlbl}$ {\scriptsize (Nm)}}
			\psfrag{umLS-st (Nm)}[Bc][Bc][2]{$u_{\mLS_\stlbl}$ {\scriptsize (Nm)}}
			\psfrag{umLS-sw (Nm)}[Bc][Bc][2]{$u_{\mLS_\swlbl}$ {\scriptsize (Nm)}}
			\psfrag{Stance Leg Torque}[Bc][Bc][1]{Stance Leg Torque}
			\psfrag{Swing Leg Torque}[Bc][Bc][1]{Swing Leg Torque}
			\resizebox{0.45\linewidth}{!}{\includegraphics{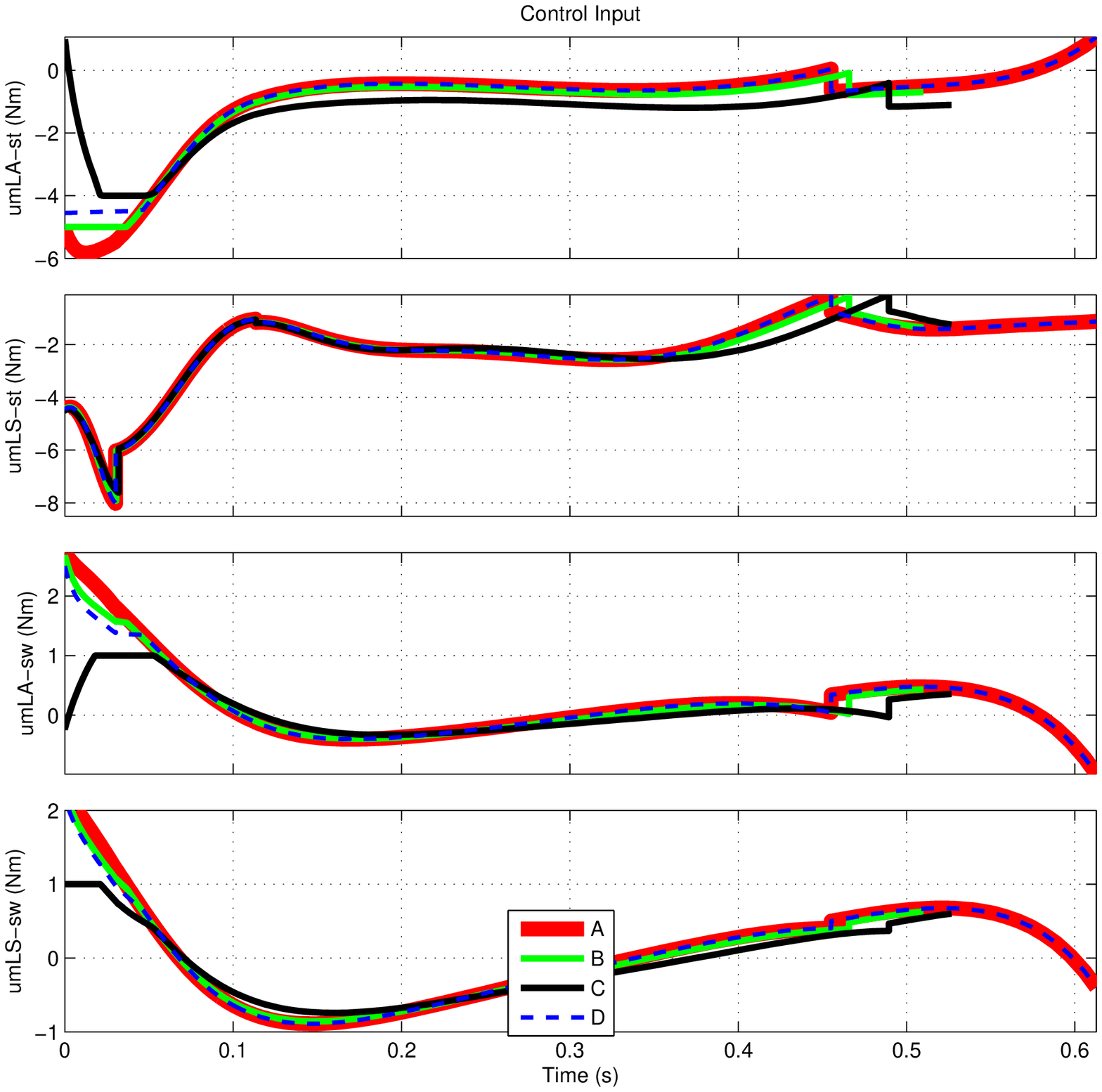}}
		}
		\subfloat[][]{
			\psfrag{Time (s)}[Bc][Bc][1.5]{Time (s)}
			\psfrag{Tracking Errors}[Bc][Bc][2]{Tracking Errors}
			\psfrag{hLA-st (deg)}[Bc][Bc][2]{$y_{\LA_\stlbl}$ {\scriptsize (deg)}}
			\psfrag{hLA-sw (deg)}[Bc][Bc][2]{$y_{\LA_\swlbl}$ {\scriptsize (deg)}}
			\psfrag{hLS-st (deg)}[Bc][Bc][2]{$y_{\LS_\stlbl}$ {\scriptsize (deg)}}
			\psfrag{hLS-sw (deg)}[Bc][Bc][2]{$y_{\LS_\swlbl}$ {\scriptsize (deg)}}
			\psfrag{Stance Leg Torque}[Bc][Bc][1]{Stance Leg Torque}
			\psfrag{Swing Leg Torque}[Bc][Bc][1]{Swing Leg Torque}
			\resizebox{0.45\linewidth}{!}{\includegraphics{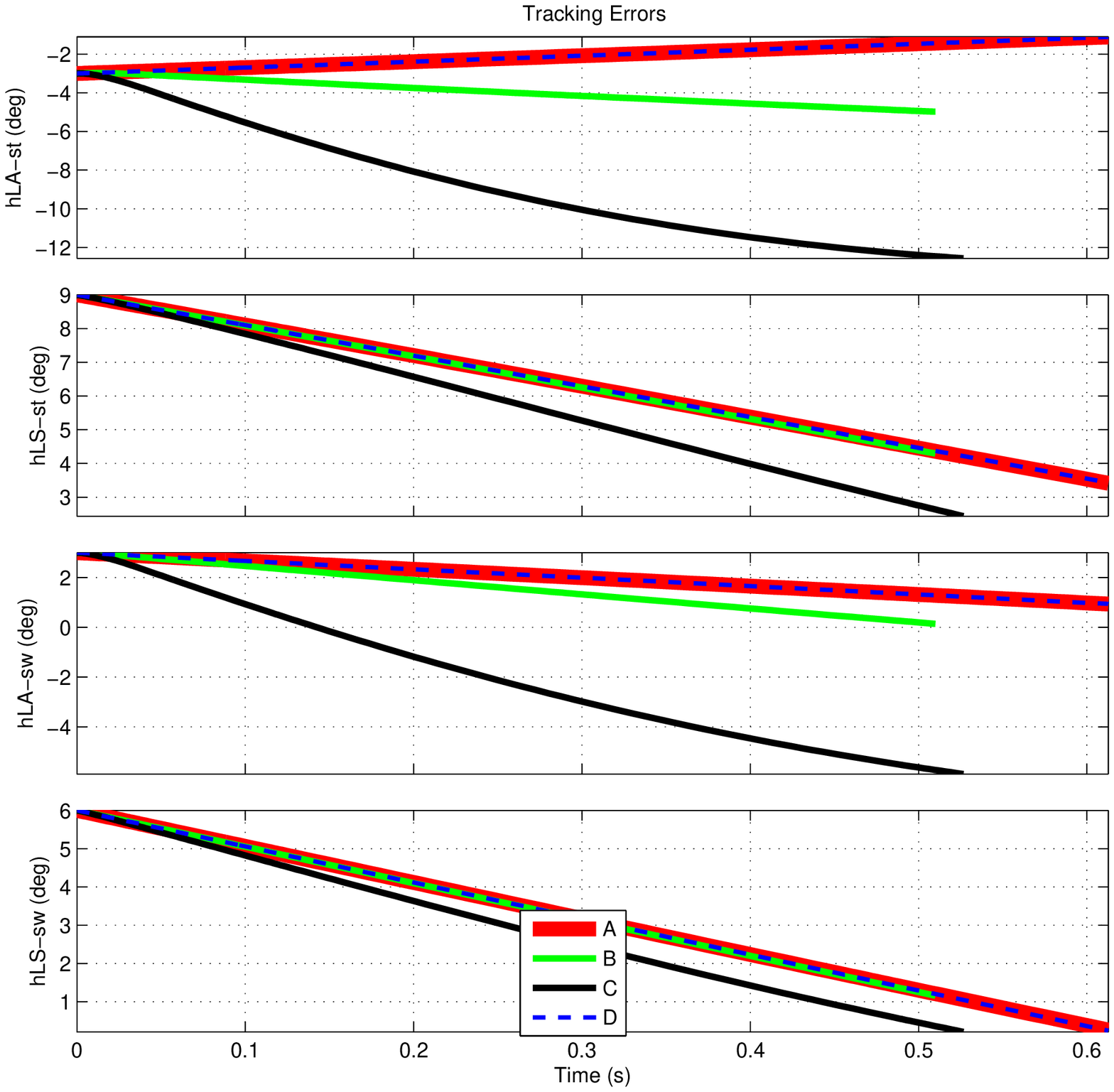}}
		}
		\caption{(a) Motor torques for the stance (top two figures) and swing legs (bottom two figures), and (b) Corresponding errors in tracking the output \eqref{eqn:robotOutputFcns}, based on the numerical simulations described in Section \ref{sec:simulation}.  Each figure depicts the results for four different cases of input saturation bounds. As can be observed in the tracking error plots, Case C leads to instability in the walking gait.}
		\label{fig:sim_torques}
	\end{figure*}

	Simulations of a representative walking step with the controller \eqref{eqn:cvxWithHardSaturations} were run for each of Cases A-D; the corresponding RES-CLF $\Ve$ and its time derivative are presented in Figure \ref{fig:sim_V_dV}, and the resulting input torques and tracking errors are illustrated in Figure \ref{fig:sim_torques}. The saturation effects are most visible in the stance leg angle plots; as expected, more restrictive saturation bounds result in increased tracking error. However, we observe that the degradation in performance is gradual. 
Note that case C leads to instability in the walking gait, as is evidenced in the plots of the tracking error in Figure \ref{fig:sim_torques} as well as the Lyapunov function in Figure \ref{fig:sim_V_dV}.
	
\begin{figure}
		\centering
		\psfrag{qT (deg)}[Bc][Bc][2]{$q_\Tor$ (deg)}
		\psfrag{dqT (deg/s)}[Bc][Bc][2]{$\dot{q}_\Tor$ (deg/s)}
		\psfrag{I}[Bc][Bc][1.5]{I}
		\psfrag{II}[Bc][Bc][1.5]{II}
		\psfrag{III}[Bc][Bc][1.5]{III}
		\psfrag{IV}[Bc][Bc][1.5]{IV}
		\resizebox{1.0\linewidth}{!}{\includegraphics{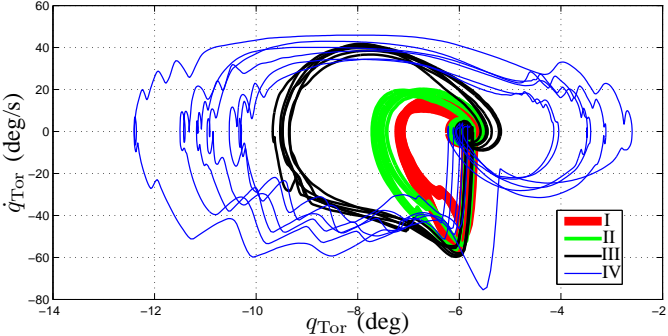}}
		\caption{Phase portrait of the torso angle for four different cases of input saturation bounds: Case I corresponds to $-8 \le u_\mLA \le 8$, $-12 \le u_\mLS \le 12$, Case II corresponds to $-6 \le u_\mLA \le 4$, $-6.5 \le u_\mLS \le 4$, Case III corresponds to $-4 \le u_\mLA \le 2$, $-6 \le u_\mLA \le 3$, and Case IV corresponds to a dynamic torque bound that is a function of the state of the robot.  Observe that stricter saturations result in (gradual) deterioration in tracking, as evidenced by deviations of the limit cycle from the nominal orbit.}
		\label{fig:sim_qT_phase_portrait}
	\end{figure}

	To illustrate the effect of saturation on the walking limit cycle, we also carry out simulations on the complex model of \mabel.  We use the controller given by \eqref{eqn:cvxWithHardSaturations} in closed-loop and analyze the phase portrait of the torso angle, subject to several different saturation values.  Figure \ref{fig:sim_qT_phase_portrait} illustrates the torso phase portrait for $15$ steps of walking, and we observe that stricter saturations result in (gradual) deterioration in tracking, as evidenced by deviations of the limit cycle from the nominal orbit. The saturation values used here differ from those used in the simulations described in the first part of this section, since the complex model significantly differs from the simple model and the required torques for walking are different.  However, the saturations are notationally similar, with the saturations becoming more restrictive as we go from Case I to Case IV (as described in the caption of Figure \ref{fig:sim_qT_phase_portrait}).

\subsection{Experimental results}
\label{sec:experiment}
	
	Motivated by the favorable numerical simulation results, we proceed to test the controller experimentally on \mabel. Experimental implementation of the CLF controller at real-time speeds is a challenging task, since it requires computation of the system dynamics \eqref{eqn:fgsystemRepeated}, the Lie derivatives of the output \eqref{eqn:robotOutputFcns}, and the CLF controller terms \eqref{eqn:psiDefnsForExperiment}, as well as the solving of a convex optimization problem. In order to meet hard real-time constraints of $1$ kHz, these computations must be completed in less than $1$ ms.  By employing the custom-code generation method CVXGEN \cite{MaBo2012} for solving constrained quadratic programs, we are able to solve the optimization problem in a few hundred microseconds and meet the $1$ kHz update requirement, making experimental implementation feasible.
	
\begin{figure}
	\centering
	\psfrag{Time (s)}[Bc][Bc][1.5]{Time (s)}
	
	\psfrag{umLA (Nm)}[Bc][Bc][2]{$u_\mLA$ {\scriptsize (Nm)}}
	\psfrag{umLS (Nm)}[Bc][Bc][2]{$u_\mLS$ {\scriptsize (Nm)}}
	\psfrag{Stance}[Bc][Bc][1]{Stance}
	\psfrag{Swing}[Bc][Bc][1]{Swing}
\resizebox{1.0\linewidth}{!}{\includegraphics{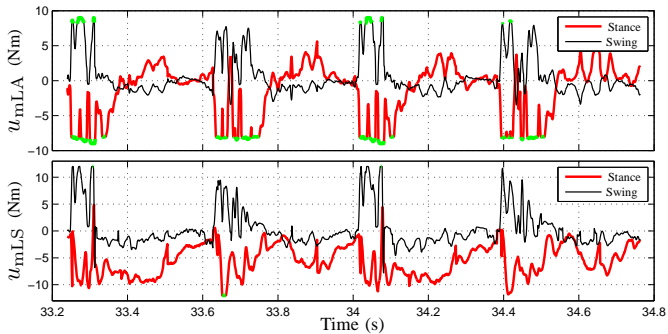}}
		\caption{Motor torques for the stance and swing legs for $4$ consecutive steps of walking with the CLF controller with convex optimization and soft constraints on saturations.  The convex optimization was tasked to enforce the magnitude of the LA and LS motor torques to be within $8$ Nm and $12$ Nm respectively.  The green highlights on the plots indicate regions where the user-specified torque bound was exceeded by the convex optimization.}
		\label{fig:CLF_cvxoptim_with_penalty_torques}
	\end{figure}

\begin{figure}
	\centering
	\psfrag{Time (s)}[Bc][Bc][1.5]{Time (s)}
	\psfrag{V}[Bc][Bc][2]{$V$}
	\psfrag{Vdot}[Bc][Bc][2]{$\dot{V}$}
\resizebox{1.0\linewidth}{!}{\includegraphics{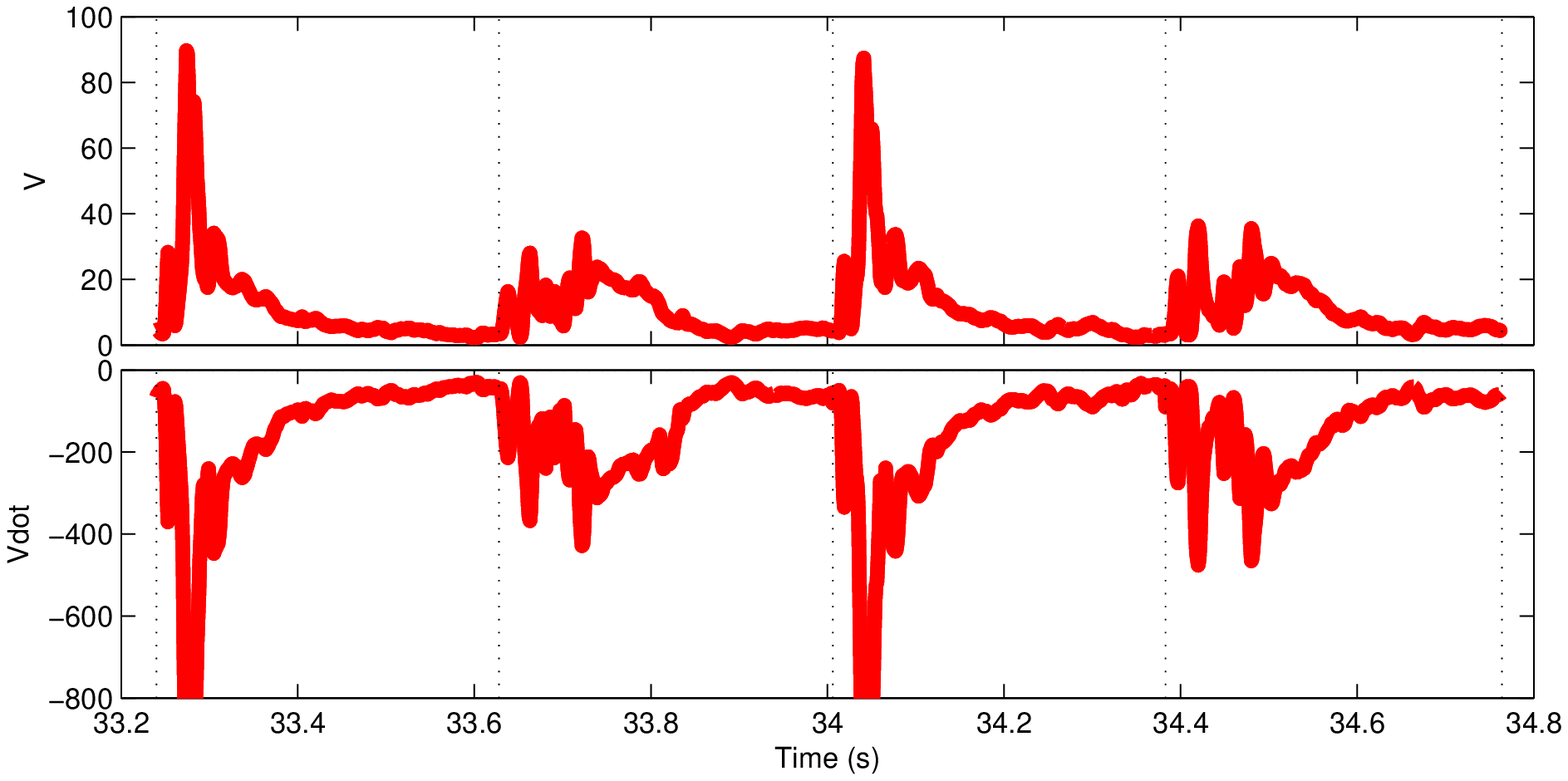}}
		\caption{Lyapunov function and its time-derivative for $4$ consecutive steps of walking with the CLF controller with convex optimization and soft constraints on saturations.}
		\label{fig:CLF_cvxoptim_with_penalty_V_Vdot}
	\end{figure}

	Two walking experiments were performed with \mabel. The first experiment employed the control method presented in \eqref{eqn:cvxWithSoftSaturations}, implementing soft bounds on the control torques with $p_1 = 50$, $p_2 = 75$, and $u_{min}, u_{max}$ chosen such that $-8 \le u_\mLA \le 8$, $-12 \le u_\mLS \le 12$.  The controller executed $169$ steps of stable walking, before the experiment was terminated by the operator.  Figure \ref{fig:CLF_cvxoptim_with_penalty_torques} illustrates the resulting torques. Note that the section of the plots highlighted in green illustrate the points at which the user-defined soft bounds on control inputs are relaxed, as the controller trades off strict adherence to the control bounds for better performance with regards to the CLF bound.  Figure \ref{fig:CLF_cvxoptim_with_penalty_V_Vdot} illustrates the Lyapunov function and its time derivative for this experiment.
	
\begin{figure}
\centering
\resizebox{1.0\linewidth}{!}{\includegraphics{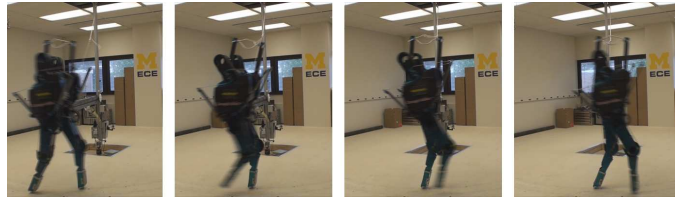}}
\caption{A photo sequence depicting one representative step from the second experiment described in Section \ref{sec:experiment}.}
\label{fig:gaitTile}
\end{figure}

\begin{figure}
	\centering
	\psfrag{Time (s)}[Bc][Bc][1.5]{Time (s)}
	
	\psfrag{umLA (Nm)}[Bc][Bc][2]{$u_\mLA$ {\scriptsize (Nm)}}
	\psfrag{umLS (Nm)}[Bc][Bc][2]{$u_\mLS$ {\scriptsize (Nm)}}
	\psfrag{Stance}[Bc][Bc][1]{Stance}
	\psfrag{Swing}[Bc][Bc][1]{Swing}
\resizebox{1.0\linewidth}{!}{\includegraphics{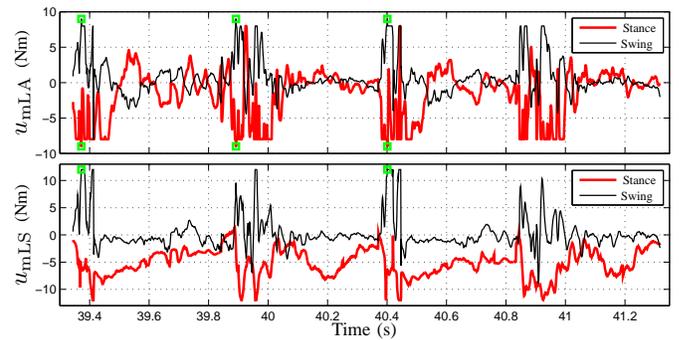}}
		\caption{Motor torques for the stance and swing legs for $4$ consecutive steps of walking with the CLF controller with convex optimization and hard constraints on saturations.  The convex optimization was tasked to enforce the magnitude of the LA and LS motor torques to be within $8$ Nm and $12$ Nm respectively.  The green square markers on the plots indicate the time instances at which the user-specified torque bound was exceeded by the convex optimization.  This occurs when the convex optimization fails to converge within the maximum number of allowed iterations, a limit required to ensure the hard real-time constraints are met for experimental implementation.}
		\label{fig:CLF_cvxoptim_without_penalty_torques}
	\end{figure}

\begin{figure}
	\centering
	\psfrag{Time (s)}[Bc][Bc][1.5]{Time (s)}
	\psfrag{V}[Bc][Bc][2]{$V$}
	\psfrag{Vdot}[Bc][Bc][2]{$\dot{V}$}
\resizebox{1.0\linewidth}{!}{\includegraphics{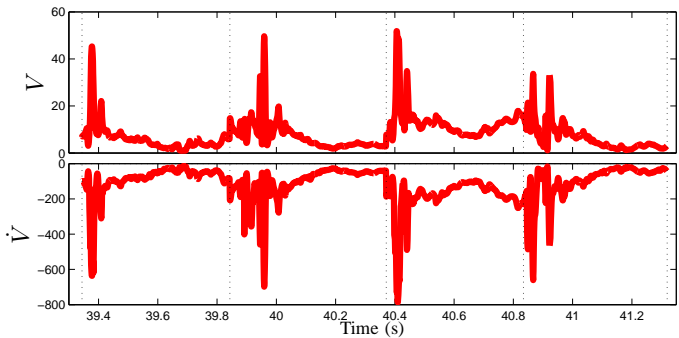}}
		\caption{Lyapunov function and its time-derivative for $4$ consecutive steps of walking with the CLF controller with convex optimization and hard constraints on saturations.}
		\label{fig:CLF_cvxoptim_without_penalty_V_Vdot}
	\end{figure}

	In the second experiment, we implemented the CLF controller with hard control saturation, as in \eqref{eqn:cvxWithHardSaturations}, with the same CLF-bound penalty as in the first experiment (i.e., $p_1 = 50$) as well as the same choices for $u_{min}$ and $u_{max}$. This experiment resulted in $70$ steps of walking for \mabel\ and is portrayed in the video in \cite{YouTubeExpCVX}. (A photo sequence depicting one representative step is also shown in Figure \ref{fig:gaitTile}.) Figure \ref{fig:CLF_cvxoptim_without_penalty_torques} illustrates the resultant control torques; we observe that the user-specified control bounds are respected, as evidenced by the flattened control signals at the boundary areas.  Note that the green squares on the plot depict the time instances at which control bounds are not met, which occur at moments in which the convex optimization algorithm is not able to converge within the specified time constraints. These occurrences are isolated and have no affect on the experimental system since a motor is not able to respond to them.  Figure \ref{fig:CLF_cvxoptim_without_penalty_V_Vdot} illustrates the Lyapunov function and its time derivative for this experiment.

%% file: sections/conclusion.tex
\section{Conclusion}
\label{sec:conclusion}
		We have presented an alternative method for implementing the pointwise min-norm CLF controller described in \eqref{eqn:minnorm} in a manner that more appropriately handles input saturations, whether those user-defined constraints are strict or soft.  Numerical simulation as well as experimental implementation has demonstrated that these control methods can be very useful in practice, even in systems which require a high real-time control update rate. This method has great potential for effectively dealing with saturations in a variety of contexts, such as power-limited systems which could progressively lower user-defined torque saturations as the battery charge decreases, thereby prolonging the last bit of battery charge while allowing system performance to gracefully degrade. In addition to dynamic torque saturation, we also note that this approach provides a method for incorporating a whole family of user-defined constraints into the online calculation of controller effort for the types of systems described here.